%% file: main.tex
\documentclass[a4paper,11pt]{article}
\pdfoutput=1 

\usepackage{jcappub} 
\usepackage[]{natbib}
\usepackage[T1]{fontenc} 
\usepackage[switch]{lineno}
\usepackage[normalem]{ulem}

\definecolor{pink}{rgb}{0.89, 0.44, 0.48}

\title{\boldmath First all-flavor search for transient neutrino emission using 3-years of IceCube DeepCore data}

\usepackage{graphicx}

\usepackage[switch]{lineno}

\emailAdd{analysis@icecube.wisc.edu}

\abstract{Since the discovery of a flux of high-energy astrophysical neutrinos, searches for their origins have focused primarily at TeV-PeV energies. Compared to sub-TeV searches, high-energy searches benefit from an increase in the neutrino cross section, improved angular resolution on the neutrino direction, and a reduced background from atmospheric neutrinos and muons. However, the focus on high energy does not preclude the existence of sub-TeV neutrino emission where IceCube retains sensitivity. Here we present the first all-flavor search from IceCube for transient emission of low-energy neutrinos, focusing on the energy region of 5.6-100\,GeV using three years of data obtained with the IceCube-DeepCore detector. We find no evidence of transient neutrino emission in the data, thus leading to a constraint on the volumetric rate of astrophysical transient sources in the range of $\sim 705-2301\, \text{Gpc}^{-3}\, \text{yr}^{-1}$ for sources following a subphotospheric energy spectrum with a mean energy of 100\,GeV and a bolometric energy of $10^{52}$\,erg.
}

\include{Authors}

\begin{document}
\maketitle
\flushbottom

%
\section{Introduction}
\label{sec:intro}
The past few years have seen the advent of high-energy neutrino astronomy, with the identification and characterization of a diffuse astrophysical flux of neutrinos at TeV energies and above~\citep{Aartsen:2016xlq, Aartsen:2013jdh,Aartsen:2014gkd}, as well as the first observation of an extragalactic source of such high energy neutrinos~\citep{IceCube:2018dnn, IceCube:2018cha}. Neutrinos nevertheless remain some of the most underutilized of the astronomical particle messengers. Their production is theorized in a wide range of astrophysical scenarios, and may be the only observable signatures of processes in dense regions of the Universe. The IceCube Neutrino Observatory (IceCube) is composed of a surface array as well as an in-ice array consisting of $\sim$1 cubic kilometer of instrumented ice located at the South Pole~\citep{Aartsen:2016nxy}. Most IceCube searches for astrophysical neutrino emission are optimized for TeV-PeV energies, relying on the enhanced effective volume (due to the increased cross sections as well as the longer propagation distances of the secondary particles) of the detector at these high energies and the steeply falling spectrum of atmospheric neutrino and muon backgrounds. In the context of considering a `stacking' analysis of sources, a 10-year data study by IceCube finds evidence for neutrino emission in a set of 4 objects, including TXS0506+056~\citep{Aartsen:2019fau}.

Searches for astrophysical sources of neutrinos at sub-TeV energies have been carried out with Super-Kamiokande~\citep{Thrane:2009tw, Abe:2018mic, Super-Kamiokande:2021dav}, the Baksan Underground Scintillation Telescope \citep{Boliev:2021log}, as well as with IceCube data~\citep{IceCube:2021ddq, Aartsen:2015eai}. The analyses have been predominantly focused on using only $\nu_\mu$, and the untriggered `tabula rasa’ search in Super-Kamiokande tested for power-law spectra. All searches suffer from significantly larger background event rates compared to the high-energy searches due to the steeper power laws that characterize atmospheric background spectra. However, they remain a promising avenue for the discovery of transient emitters, which are in general not expected to have power law spectra~(\cite{longair_1994}, see chapter 21). Scenarios that predict flaring neutrino emission which peak in the 10-100\,GeV range along with a suppression of emission at higher energies are of particular interest. This may occur due to the subphotospheric emission of neutrinos from gamma-ray bursts (GRBs), wherein the jet is dissipated by inelastic collisions and quasithermal neutrinos are produced via hadronuclear reactions~\citep{Murase:2013hh}. Detecting sub-TeV neutrinos from GRBs can thus delineate the subphotospheric scenario from the more classical case, in which gamma-rays are attributed to synchrotron emission from nonthermal electrons accelerated at internal shocks~\citep{Toma:2010xw}. Neutrino emission that peaks at $\sim$GeV energies is also naturally expected in core collapse supernovae in which the jets are not energetic enough to break through the stellar envelope and gamma-ray emission is thus `choked' off~\citep{Ando:2005xi}. The observed correlation between long duration GRBs and a small fraction of core-collapse supernovae~\citep{Modjaz:2011bm} hints that a larger number of core collapse supernovae may develop such choked jets. It has been suggested~\citep{Taboada:2010xf} that neutrino emission from such events occurring  within $\sim$10\,Mpc should be detectable by the IceCube-DeepCore sub-array \citep{, IceCube:2021ddq}.

While subphotospheric GRBs and choked jets provide a class of interesting neutrino emission candidates, it is not necessary to have a specific targeted model in order to observe neutrino emission from  yet unknown astrophysical phenomena, as evidenced by the serendipitous detection of astrophysical neutrinos above $\mathcal{O}(10)$\,TeV by IceCube~\cite{Aartsen:2013jdh}. As such, we model potential neutrino sources as a comprehensive Fermi-Dirac-like emission spectra with  effective temperatures that are peaked at two representative energies, 20\,GeV and 100\,GeV, which fall within the 5.6-100\,GeV energy range covered by the event selection, discussed further in Sec.\@ \ref{sec:DetectorEventSample}. The particular form of the emission is a) inspired by subphotospheric emission models~\citep{Murase:2013hh}, and b) corresponds to neutrino emission that would not produce a large contribution at energies greater than $1$\,TeV.


This work presents the search for generic transient emitters of neutrinos, with long duration gamma-ray bursts used as proxy for potential sources, in the energy range of 5.6-100\,GeV with three years of IceCube-DeepCore data and marks the first use of an all-flavor sample of neutrinos at these low energies for such purposes. While the inferior angular resolution of, and limited effective volume for, $\nu_e$ (+ $\bar{\nu}_e$) and $\nu_{\tau}$ (+ $\bar{\nu}_{\tau}$) events at high energies (TeV) discourage their use in high-energy searches~\citep{Aartsen:2019fau}, in low-energy searches targeting transient emission, they form an additional signal component that enhances the potential for discovery. It is worth noting that the energy sensitivity for this analysis is above the $\sim3.5$\,GeV threshold for charged current $\nu_\tau$ interactions, so this is an all-flavor analysis with regards to the neutral current and charged current channels.

\section{Detector and event sample}
\label{sec:DetectorEventSample}
The IceCube Neutrino Observatory consists of roughly 1 cubic kilometer of instrumented glacial ice located at the South Pole~\citep{Aartsen:2016nxy}. The detector contains 86 vertical strings each composed of 60 digital optical modules (DOMs) including a downward facing 10-inch photomultiplier tube. The DOMs are deployed from 1.45\,km to 2.45\,km below the surface of the ice mostly in a hexagonal pattern. The largest part of the detector, the in-ice array, is optimized for neutrinos at $\sim$ TeV -- PeV energies and has some sensitivity to the detection and analysis of neutrinos in the $\mathcal{O}(100)$\,GeV region. IceCube-DeepCore is the more densely instrumented region located approximately in the middle of the detector~\citep{Collaboration:2011ym}. It consists of the most central standard IceCube strings as well as 8 specialized low-energy strings whose DOMs have a higher quantum efficiency than the regular IceCube DOMs. Hence, this part of the detector is sensitive to neutrinos at energies down to $\sim$5\,GeV.

\begin{table}[b]
\centering
\caption{Relative fractions of event types from Monte Carlo simulation for the GRECO dataset for each neutrino flavor (assuming an atmospheric flux), atmospheric muons, and accidental events caused purely by detector noise \citep{Aartsen:2019tjl}. This is the event distribution used for background in our search.}
\label{tab:GRECO_ratio}
\begin{tabular}{ c c }
\hline \hline
 \bf{Event Type} & \bf{Relative Fraction (\%)} \\  \hline
 $\nu_\mu + \bar{\nu}_\mu$ &  58.68  \\ \hline
 $\nu_e + \bar{\nu}_e$ &  28.21 \\ \hline
 $\nu_\tau + \bar{\nu}_\tau$ & 4.43  \\ \hline
 $\mu$ & 8.68  \\ \hline
 pure-noise & <0.01\\ \hline \hline
\end{tabular}
\end{table}

\begin{figure}
	\center
	\includegraphics[width=0.7\linewidth]{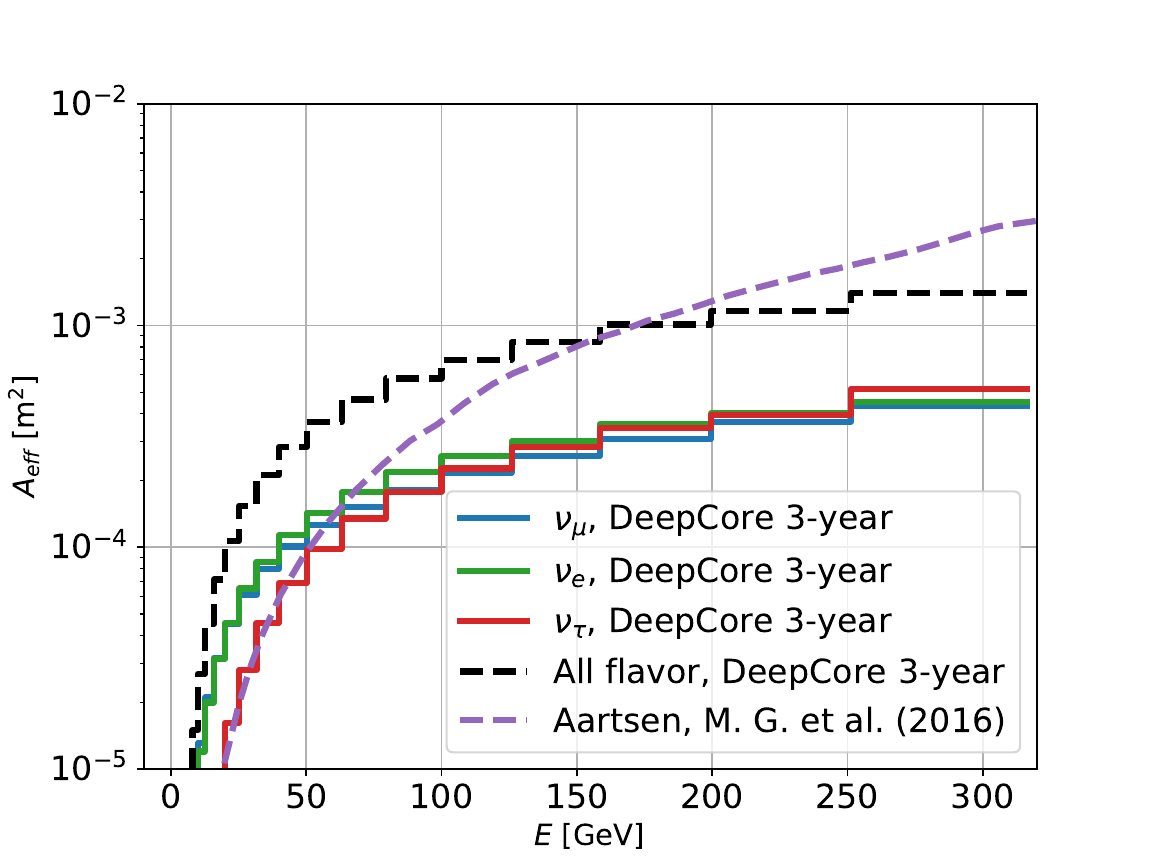}
	\caption{All sky (4$\pi$\,sr) average neutrino effective area for the GRECO event selection. The effective area of the previous low-energy transient analysis from IceCube \citep{Aartsen:2015eai}, which was only $\nu_\mu$, is indicated by the purple dashed line.}
	\label{fig: Effective_area}
\end{figure}

In this work, we use the GRECO (GeV Reconstructed Events with Containment for Oscillation) event selection. This selection was originally developed for the tau neutrino appearance analysis which is labelled as `analysis $\mathcal{A}$' in previous work~\citep{Aartsen:2019tjl}, specifically the `L7' sample which includes events with a reconstructed energy $>$56\,GeV. The three years of data are obtained from April 2012 through May 2015 and have a livetime of $\sim$2.756 years (1006 days) and an average rate of 0.87\,mHz. The effective area for the event sample is shown in Figure~\ref{fig: Effective_area}, where also the individual contributions for the different neutrino flavors are indicated. This is compared to the effective area for a previous low-energy transient point-source search performed with IceCube data \citep{Aartsen:2015eai}. The background to any potential transient neutrino signal is a combination of atmospheric neutrinos, atmospheric muons, and PMT/DOM glass/electronics noise events that manage to trigger the array and pass the event selection. For the GRECO data the relative contributions of each of the major background types are shown in Table~\ref{tab:GRECO_ratio}.

\section{Analysis method}
\label{sec:AnalysisMetod}
This analysis consists of two parts: a kernel density estimation (KDE) used on the event times with the purpose of selecting time windows with event densities that are larger than a threshold defined to exclude background. This is followed by a maximum likelihood analysis that tests for spatial clustering of the events observed close together in time. 

While establishing the performance of the analysis, time scrambled data is used for the background estimation. This means that we use experimental data obtained with IceCube-DeepCore but assign random observation times within the run-time of the detector for the individual events. Using scrambled data ensures that the algorithms and parameters values are developed and selected without knowledge of the final analysis results.

\subsection{Kernel density estimation}
\label{sec:KDE}

\begin{figure}
	\center
	\includegraphics[width=0.7\linewidth]{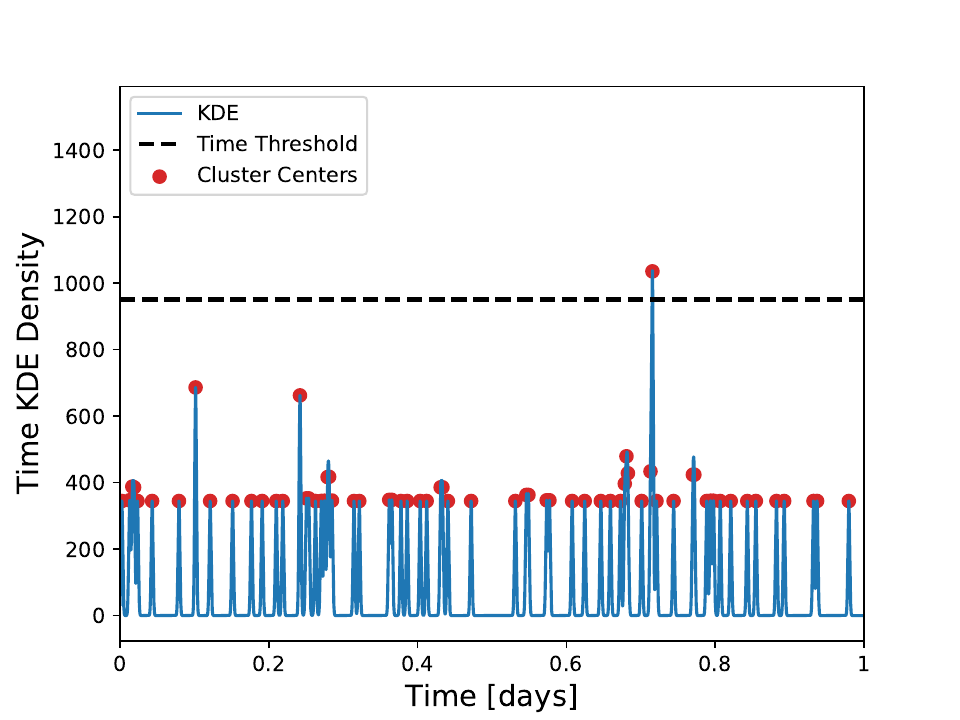}
	\caption{Example of time KDE. The x-axis is unit fraction of one day. Gaussian kernels for individual events are summed in order to compute the density of events with respect to time. No normalization is performed as this would change the scale of the y-axis for different numbers of events. }
	\label{fig: time KDE}
\end{figure}

The purpose of the time KDE is to identify the regions in time with large densities of events and thereby select the time windows of interest. Transient sources are expected to emit neutrinos primarily within a relatively short amount of time, and therefore the selected time windows with larger event densities are more likely to contain astrophysical transient emission. This way, the time KDE functions as a preselection of the most promising events.

When evaluating the KDE, a Gaussian kernel defined by 
\begin{linenomath}
\begin{equation} \label{eq: kernel} 
    K(t,t_n,\sigma) = \frac{1}{\sqrt{2\pi}\sigma} \exp \left(-\frac{|t-t_n|^2}{2\sigma^2} \right),
\end{equation}
\end{linenomath}
is created around the time of each event with a bandwidth $\sigma$ that should correspond to the neutrino emission time of the astrophysical transient source class under consideration. The value of $t-t_n$ is the difference between the kernel center and a given event. Since the width of the time distribution of $\mathcal{O}(5-100)$\,GeV neutrinos emitted from any sources has yet to be determined, we assume a bandwidth of 100\,s corresponding approximately to the gamma-ray emission time for long duration gamma-ray bursts~\citep{Harmon:2004wm}.

\begin{figure*}[t]
	\center
	\includegraphics[width=0.49\linewidth]{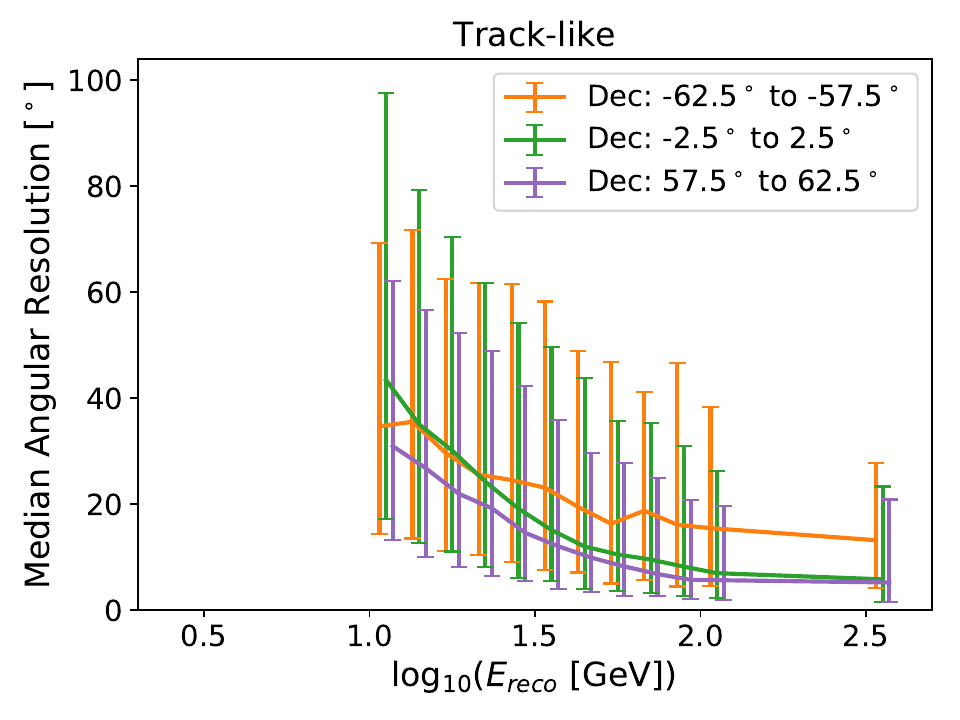}
	\includegraphics[width=0.49\linewidth]{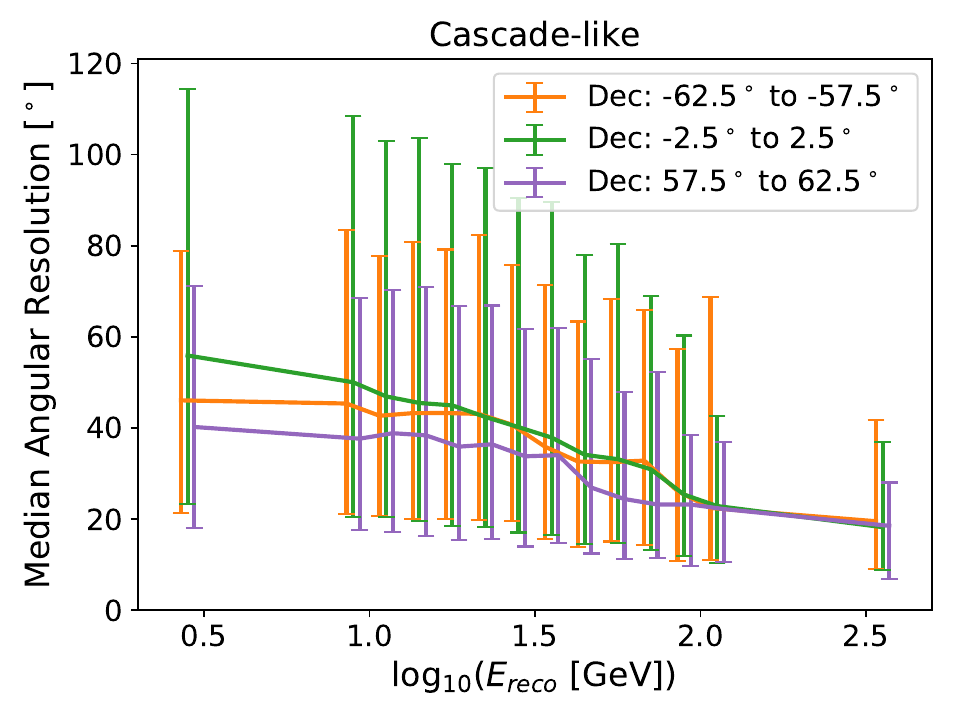}
	\caption{Median angular resolutions, i.e., opening angles between true and reconstructed event directions, used for event angular resolution splines for three example declinations. Left panel shows the result for tracks, while the right panel shows the result for cascades. The error bars indicate the 68\% central intervals and are located at the energy bin centers. The asymmetry in the error bars illustrates that there is a `longer tail' of larger angular resolutions than smaller resolutions. Orange and purple error bars have been shifted slightly towards smaller and larger values, respectively, in order to avoid overlapping. }
	\label{fig: spline}
\end{figure*}

The sum of the individual contributions from the kernels is evaluated by
\begin{linenomath}
\begin{equation} \label{eq: R_KDE}
    R_{KDE}(t,\sigma) = \sum _{n=1}^{N} K(t,t_n,\sigma),
\end{equation}
\end{linenomath}
where $N$ is the total number of events (kernels). All maxima in time, denoted as {\sl cluster centers}, are found, and the maxima above a chosen threshold on the value of $R_{KDE}$ are selected. A time window with a potential transient (TWPT) of $\pm$3 bandwidths (600\,s) is created around each cluster center passing the threshold. The threshold is chosen to allow a given average number of cluster centers per year to pass. One time window is created around each, resulting in an expected average number of time windows per year (this parameter will be denoted as $N_{TWPT}$). In this case, an $R_{KDE}$-threshold of 905.80 is chosen corresponding to $N_{TWPT}=$100. A relatively small number of clusters passing the threshold is chosen because a larger number provides no appreciable improvements in the sensitivity. Figure~\ref{fig: time KDE} shows an example for one day of scrambled data.

\subsection{Likelihood Maximization}
\label{sec:MaximumLikelihood}

A maximum likelihood analysis is performed on the events within each selected time window created based on the time KDE. The likelihood is given by
\begin{linenomath}
\begin{equation}
    \mathcal{L}(n_s) = \frac{(n_s + n_b)^N}{N!}e^{-(n_s + n_b)} \prod_{i=1}^N \left( \frac{n_s \mathcal{S}_i}{n_s + n_b} + \frac{n_b \mathcal{B}_i}{n_s + n_b} \right),
\end{equation}
\end{linenomath}
where the first two factors describe the Poisson distributed probability of observing $N$ events from the source under consideration. 
The log-likelihood ratio is then given by
\begin{linenomath}
\begin{equation} \label{eq: log-likelihood ratio}
    \ln \left( \frac{\mathcal{L}(n_s)}{\mathcal{L}(0)} \right) = -n_s + \sum_{i=1}^{N} \ln \left( \frac{n_s \mathcal{S}_i}{\left< n_b \right> \mathcal{B}_i} + 1 \right).
\end{equation}
\end{linenomath}
The parameters $n_s$ and $n_b$ are the numbers of signal and background events within the given time window, while $\left< n_b \right>$ is the expected average based on the average background rate and the size of the time window. The parameter $n_s$ refers to the number of signal events being fit during the maximization. The variables $\mathcal{S}_i$ and $\mathcal{B}_i$ are the signal and background probability distribution functions (PDFs), i.e, the probabilities that the event $i$ within a time window originates from the source and the background, respectively.

The background PDF $\mathcal{B}_i$ is assumed to be constant in azimuth, but changes as a function of the zenith angle of the event $i$. The value is calculated by
\begin{linenomath}
\begin{equation}
    \mathcal{B}_i = \mathcal{P}_{i, zenith} \cdot \mathcal{P}_{i, azimuth} = \mathcal{P}_{i, zenith} \cdot \frac{1}{2\pi} ,
\end{equation}
\end{linenomath}
where $\mathcal{P}_{i, zenith}$ is found using a spline with 25 bins in the cos(zenith) distribution of events, i.e., the background PDF roughly follows the zenith distribution of atmospheric neutrinos with a small contribution from atmospheric muons, and is based on experimental data.

The signal PDF $\mathcal{S}_i$ contains spatial information about the potential sources as well as the events. The expected angular resolutions of the events are estimated from their reconstructed energies and declinations using a spline of the median angular resolution found from simulated GRECO events. In this case, the resolution refers to the opening angle between the true and reconstructed event directions, and the data have been weighted according to the expected energy spectrum for atmospheric neutrinos described by~\citep{Honda:2015fha}. Separate splines are employed for each event type (tracks and cascades) in order to account for the better angular resolutions of the tracks. In this case, tracks are defined as events with reconstructed track lengths of at least 50\,m. The splines are functions of energy and declination for the events. Bins in declination are distributed linearly in the range -85$^\circ$ to 85$^\circ$, with bin centers 5$^\circ$ apart. Bins in energy are logarithmically distributed with an additional bin at higher and/or lower energies (see Figure~\ref{fig: spline}).

The signal PDF $\mathcal{S}_i$ is defined as a Kent distribution, i.e., a probability distribution analogous to the bivariate normal distribution but normalized correctly on a sphere. Assuming circular angular errors on the events, i.e, angular resolutions in azimuth and zenith are equal, the Kent distribution takes the form
\begin{linenomath}
\begin{equation} \label{eq: signal PDF}
    \mathcal{S}_i = \frac{\kappa_i}{4\pi \sinh(\kappa_i)} \exp \left( \kappa_i \cos(|x_{\mathrm{source}}-x_i|) \right).
\end{equation}
\end{linenomath}
Here, $|x_{\mathrm{source}}-x_i|$, is the angular distance between the source and the event $i$. The parameter $\kappa_i$ is the concentration parameter for the distribution, related to the angular resolution of the events by
\begin{linenomath}
\begin{equation} \label{eq: kappa}
    \kappa_i \approx \frac{1}{\sigma_i^2},
\end{equation}
\end{linenomath}
where the median of the differences between true and reconstructed event directions are used as the angular resolution. This will be discussed more thoroughly below. The angular extent of the source is omitted since transient emitters are expected to be point sources.

In addition to the $R_{KDE}$-threshold, a threshold is applied to the number of signal events $n_s$ in order to require a minimal amount of spatial clustering (spatial correlation) between the events within a time window. In this work, the $n_s$-threshold is set to 2. 
The log-likelihood ratio defined in Equation~\ref{eq: log-likelihood ratio} is maximized with respect to the value of $n_s$ and the source direction, $x_{source}$, included in the Kent distribution. Several distinct directional maxima may exist, and therefore we perform the maximization once for each event within the time window, while using the best-fit event position as the initial guess for the position of a possible transient point source. This ensures that all relevant maxima are found, and subsequently, the largest maximum passing the $n_s$-threshold is selected.

\subsection{Performance}
\label{sec:Performance}

\begin{figure*}
	\center
	\includegraphics[width=0.49\linewidth]{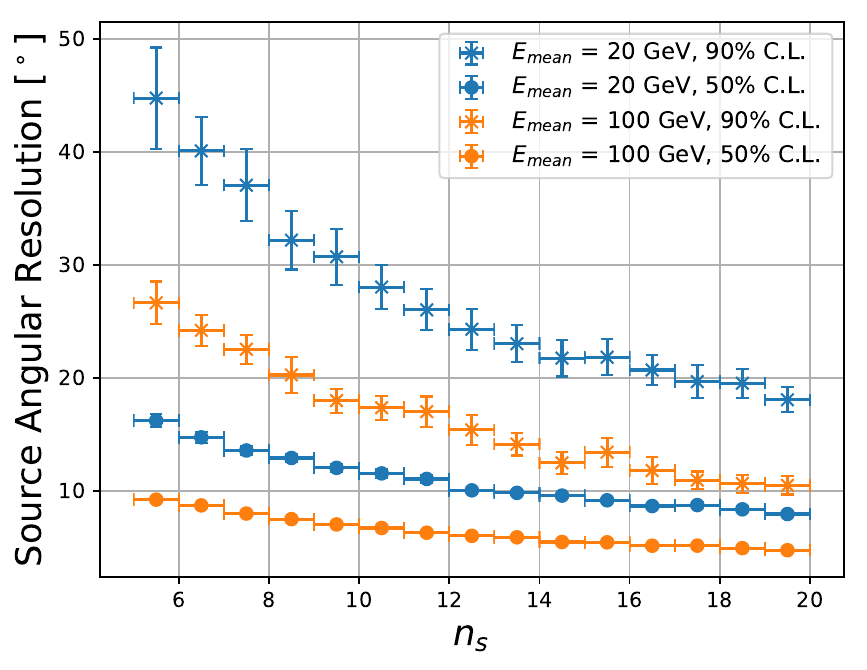}
	\includegraphics[width=0.49\linewidth]{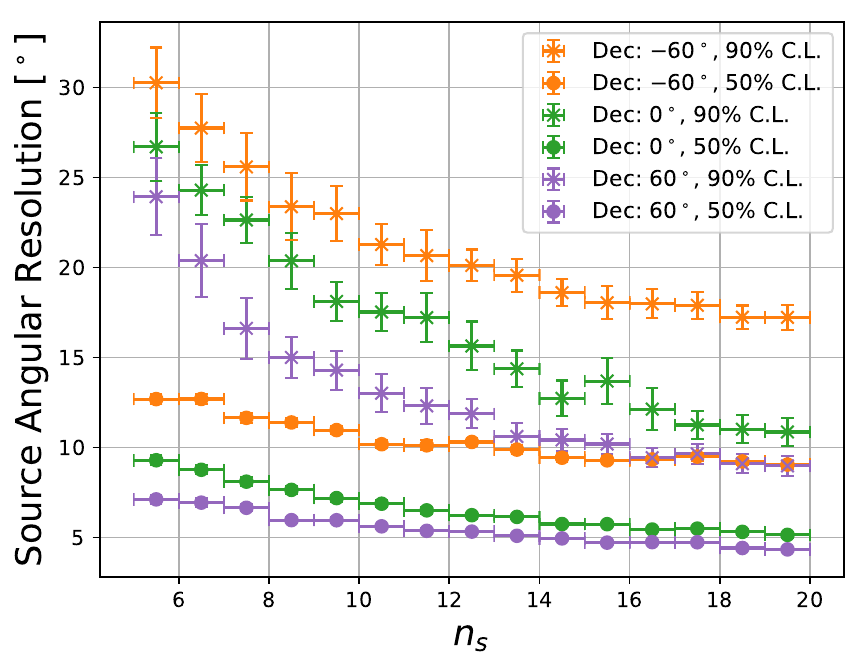}
	\caption{Uncertainty of a simulated source position as function of the number of injected neutrinos, $n_s$. Left panel: The uncertainty decreases with the neutrino mean energy for a fixed declination of 16 degrees. Right panel: Same plot but varying the declination for a fixed mean energy of 100\,GeV.}
	\label{fig: source_uncertainty}
\end{figure*}

The performance of the analysis is estimated with simulations by injecting simulated signal events into the scrambled background data. Thereby the uncertainty on the likely source position is found when maximizing the natural log of the likelihood ratio as well as the sensitivity and discovery potential for the analysis.

For the injection of signal events into the background data, the signal neutrinos are weighted according to a Dirac spectrum that mimics the shape expected from an astrophysical transient~\citep{Murase:2013hh}. The Dirac spectrum is parameterized as
\begin{linenomath}
\begin{equation}\label{eq: Dirac_fluence}
    \Phi(E) = N \cdot \frac{\left( \frac{E}{kT} \right)^2}{e^\frac{E}{kT} + 1},
\end{equation}
\end{linenomath}
where $kT$ depends on the mean energy of the neutrinos
\begin{linenomath}
\begin{equation}
    kT = \frac{E_{\mathrm{mean}}}{3.15}. 
\end{equation}
\end{linenomath}
In this work, we use mean energies of ${E_{\mathrm{mean}}=\mathrm{20}}$\,GeV and $E_{\mathrm{mean}}=100$\,GeV as benchmarks. These values serve as examples of reasonable mean energies for possible transient sources under consideration, see also Section \ref{sec:intro}. However, the analysis is not limited to, or valid only for, these specific values.

The uncertainty on the source position is defined as the difference between the true injected source position and the likely position found when performing the likelihood maximization. In this case, only the time windows actually containing the source are considered. The uncertainty is estimated and the result is shown in Figure~\ref{fig: source_uncertainty}. 
The left panel shows that the uncertainty decreases with the number of signal events $n_s$ and with the neutrino energy. The source position is generally estimated more accurately when more neutrinos from the source are detected. Higher energy neutrinos are reconstructed better, leading to the events being observed closer to the true source position. This pattern is also seen in Figure~\ref{fig: spline} for the angular resolution splines. The right panel of Figure~\ref{fig: source_uncertainty} shows that the uncertainty on the source position decreases with larger (more positive) values of the source declination, which is also a result of the event angular resolution being better at larger declinations. The angular resolutions strongly affects the signal PDF since it is being squared in Equation~\ref{eq: kappa} and because $\kappa$ is included in the exponent in Equation~\ref{eq: signal PDF}. The central median values of the angular and energy resolution are used in this transient analysis for each candidate neutrino within the examined time window(s). While the 68\% confidence intervals for the resolutions are not explicitly used, they reflect the spread of neutrino angular resolutions that are used for the pseudo-trials necessary to establish the sensitivity to an astrophysical transient source.

\begin{figure*}
	\center
	\includegraphics[width=0.49\linewidth]{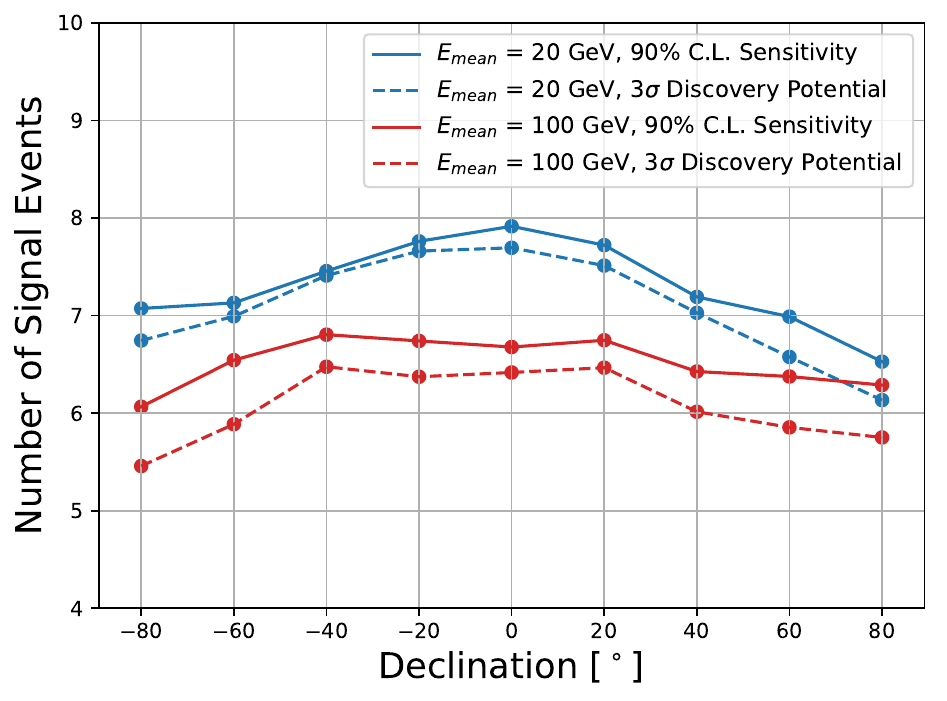}
	\includegraphics[width=0.49\linewidth]{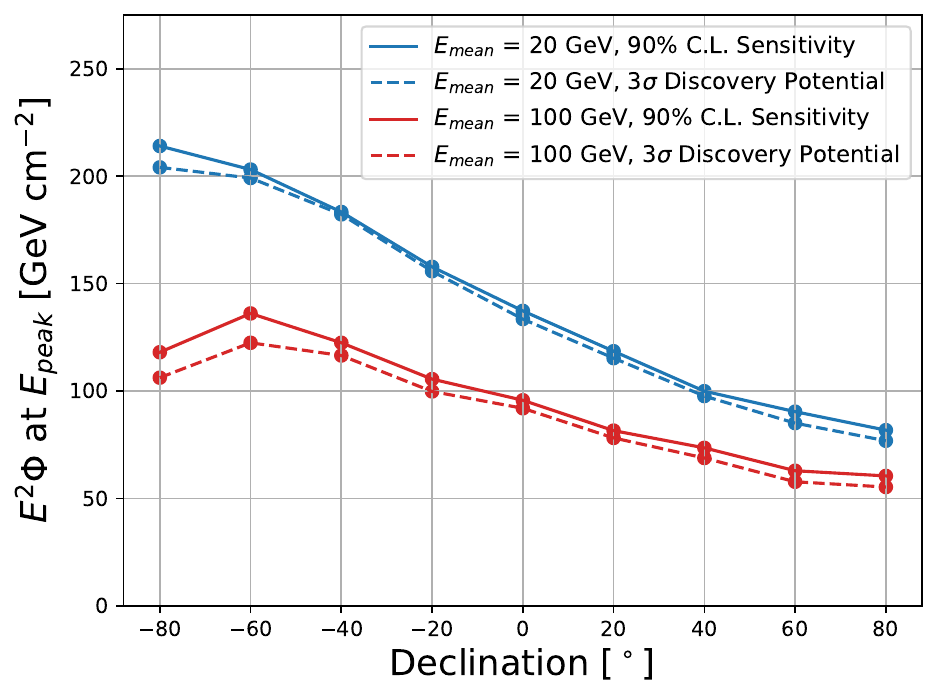}
	\caption{Left panel: Event sensitivity and discovery potential for the two different neutrino emission energies. The number of signal events on the y-axis refers to the average number of injected signal events. Right panel: The event sensitivity and discovery potential converted to fluences using the effective area for the GRECO event selection. The peak of the fluence is used as reference point for the energy (See also Section \ref{sec:UnblindResults}).}
	\label{fig: Sen_dis}
\end{figure*}

\begin{figure}
	\centering
	\includegraphics[width=0.71\linewidth, trim={0.4cm 0 0.6cm 1.4cm}, clip]{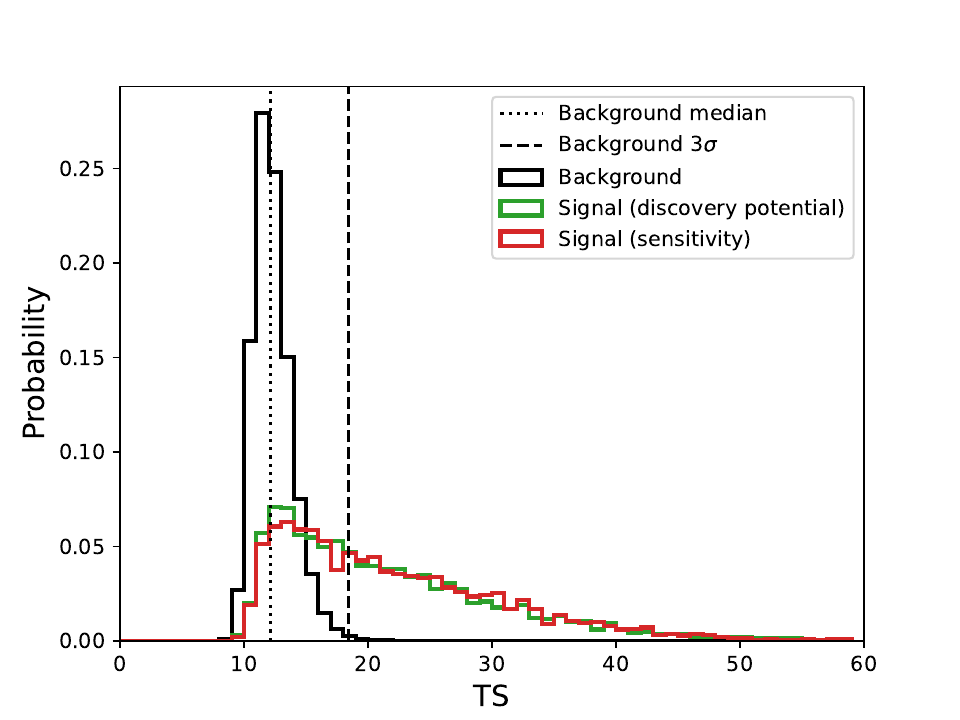}
	\caption[Test statistic distributions]{Test statistic distributions for background based on 100000 trials and two signal scenarios based on 5000 trials with the numbers of signal events corresponding to the sensitivity and discovery potential, for illustrative purposes. The signal distributions are created using the energy spectrum (\ref{eq: Dirac_fluence}) with $E_{mean}=20$\,GeV, with a source injected at a declination of 16 degrees. The sensitivity is reached for the number of injected signal events that results in 90\% of the signal test statistic distribution being greater than the median of the background, i.e., to the right of the dotted line. The discovery potential is reached when the median of the signal TS distribution is equal to the 3$\sigma$ value of the background TS distribution (dashed line).}
	\label{fig:TS_dists}
\end{figure}

In this work, we define the test statistic (TS) as the largest value of the maximized natural log of the likelihood ratios for either observed data or a background-only hypothesis. Owing to the fact that Wilks' theorem is not satisfied for this analysis we are unable to define an analytical expression for the TS distributions. Therefore, the background TS distribution used to establish the sensitivity and discovery potential for this analysis is generated from scrambled trials, where a scrambled trial corresponds to a year of time scrambled data. This motivates an estimation of a discovery potential at a 3$\sigma$ significance rather than at the more traditional 5$\sigma$ due to computational challenges. Additionally, a 3$\sigma$ discovery potential is further motivated by the possibility of using this analysis for sending transient astrophysical alerts~\citep{Smith:2012eu,Rutledge:1998hc,Aartsen:2016lmt}, in which case 3$\sigma$ is a reasonable threshold.

Figure~\ref{fig: Sen_dis} shows the 90\% sensitivity and 3$\sigma$ discovery potential for the analysis in terms of the number of observed signal events as well as fluence. The 90\% sensitivity is defined as the signal strength that results in 90\% of the scrambled trials containing signal having a value of the TS equal to or larger than the median background TS. The 3$\sigma$ discovery potential requires the median of the scrambled trials containing signal to coincide with the 3$\sigma$ background TS value. This concept is illustrated in Figure~\ref{fig:TS_dists}, where the background TS distribution is shown along with examples of TS distributions obtained with different numbers of injected signal events. The sensitivity in terms of the number of signal events is slightly worse around the horizon, since the background rate at those declinations is higher. The sensitivity in terms of the fluence is better for larger declinations due to the effective area of the detector being better in the Northern Sky.

\begin{figure*}
	\center
	\includegraphics[width=0.48\linewidth]{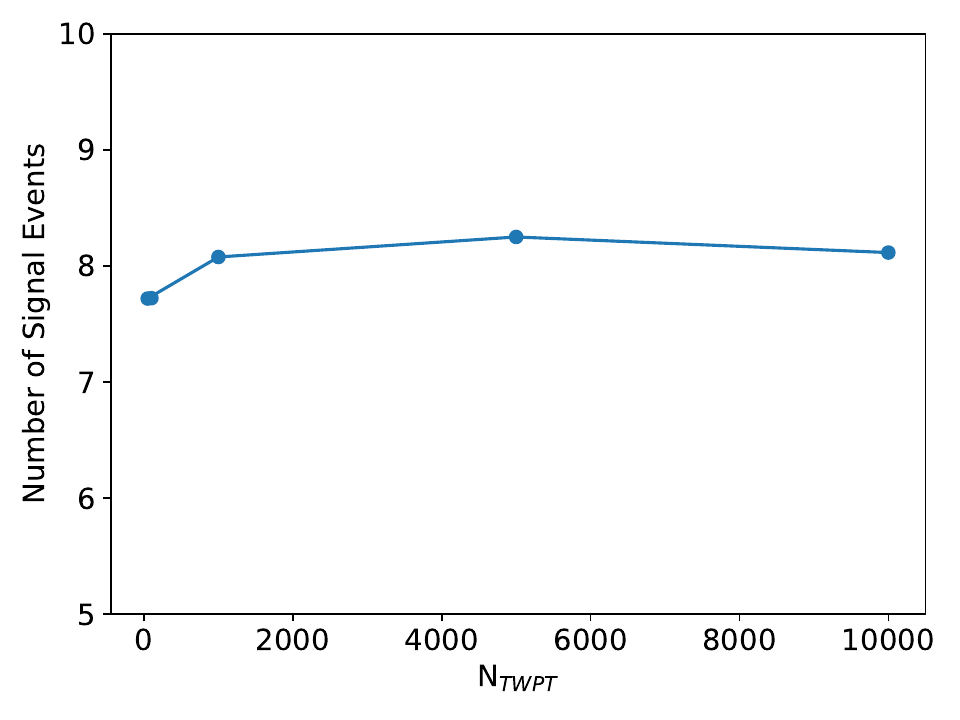}
	\includegraphics[width=0.48\linewidth]{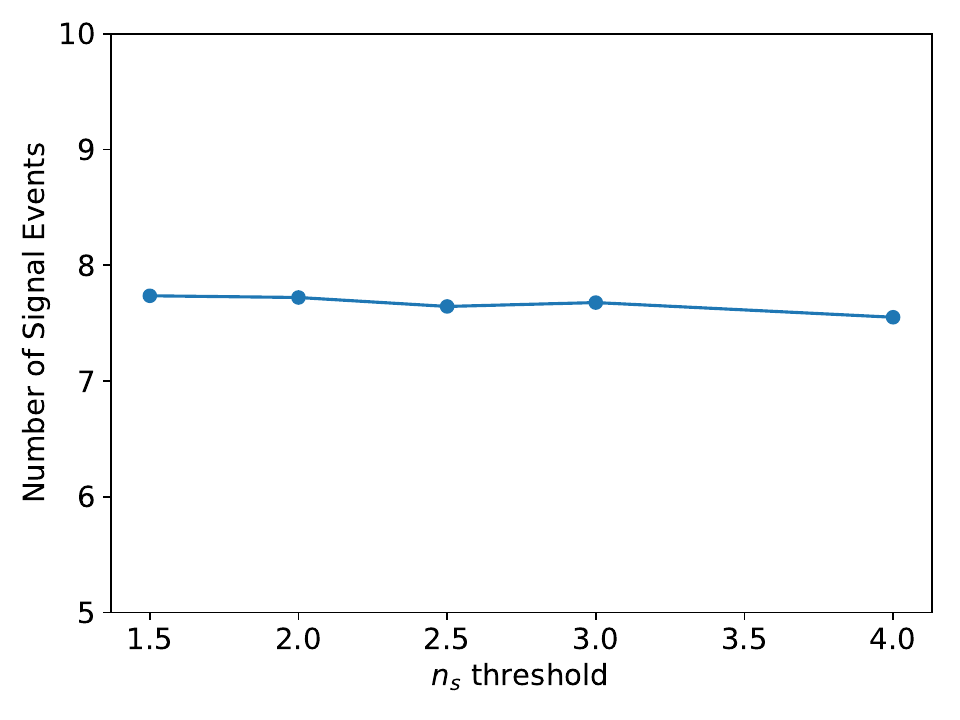}
	\caption{Both plots show the 90\% sensitivity for a source emitting neutrinos with a mean energy of \mbox{$E_{\mathrm{mean}}=20$\,GeV} located at declination of 16 degrees. The number of signal events on the y-axis refers to the average number of injected signal events. Left: The $n_s$-threshold is fixed at 2, while the KDE-threshold/average number of time windows (N$_{TWPT}$) is varied. Right: Average number of time windows is fixed at 100 per year and the $n_s$-threshold is varied.}
	\label{fig: sen_thresholds}
\end{figure*}

Figure~\ref{fig: sen_thresholds} shows that the number of signal events required to observe a transient emitter is largely constant across a range of values of N$_{TWPT}$. The final value of N$_{TWPT}=$100 was chosen to maximize the sensitivity (or minimize the required number of observed signal neutrino events), while also observing the largest number of time windows in order not to miss a potential transient neutrino emitter.
 Also, the $n_s$-threshold does not noticeably affect the sensitivity and the parameter is therefore kept fixed at 2, in order to require a minimum of spatial clustering between the events.

\begin{figure}
	\center
	\includegraphics[width=0.71\linewidth]{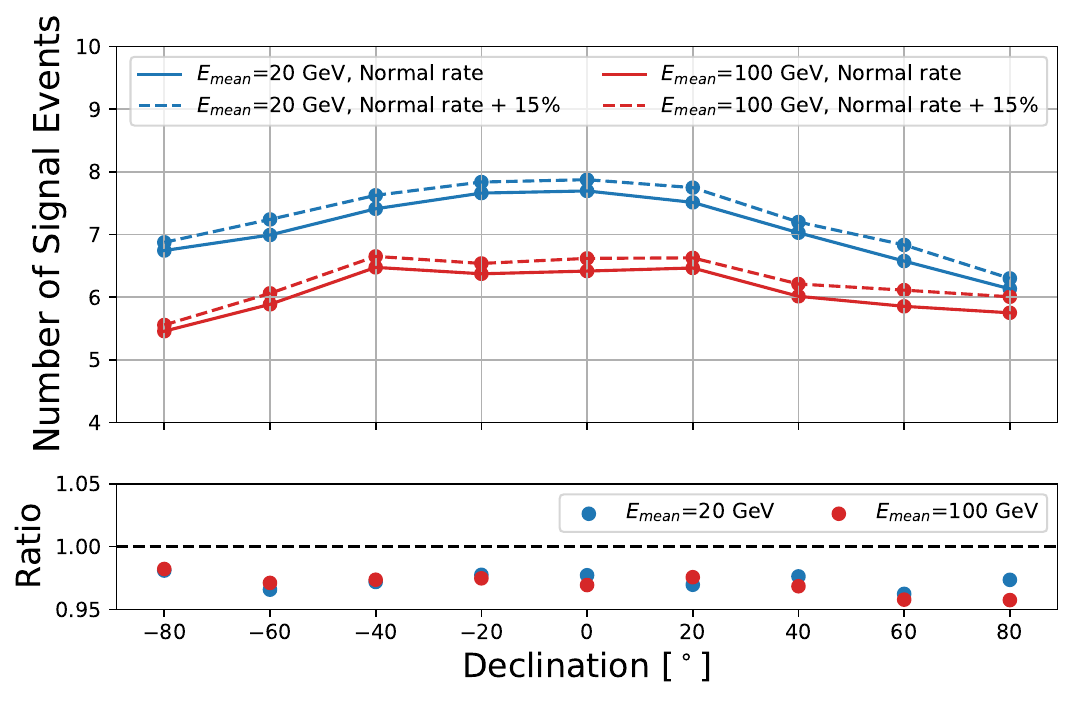}
	\caption{Comparison between the discovery potential using the actual event rate and using a rate increased by 15\%. The number of signal events on the y-axis refers to the average number of injected signal events. The red dots in the ratio plot show the ratio between the discovery potential for a source with a mean energy of 100\,GeV with a normal background rate and with an increased rate. Similarly, the blue points in the ratio plot corresponds to the ratio between the discovery potentials with normal rate and increased rate for a source with a mean energy of 20\,GeV.}
	\label{fig: Sen_dis_15}
\end{figure}

The analysis does not appreciably improve when considering shorter bandwidths; because the emission scenarios under consideration in this work have an emission duration that matches the bandwidth, and because with a background rate of 0.87\,mHz the background expectation is already so small.

The sensitivity of the analysis decreases with larger time windows (wider kernels) due to the increased atmospheric neutrino background. Ideally, all of the emitted neutrinos should be included within a single time window, which simultaneously includes the least amount of atmospheric background neutrinos. Hence, the optimal width of the time windows is comparable to the emission time of the astrophysical sources. A shorter time window would be incapable of including all, or most of, the emitted neutrinos from a source with longer emission time, therefore resulting in a significantly lower sensitivity. As such, a time window of 600\,s (which is equivalent to a duration of $\pm3$ KDE bandwidths) will make the search method developed in this work equally sensitive to sources with emission times shorter than 100\,s. While a narrower time window will exclude some background neutrinos it would also limit the analysis in terms of detectable sources (to bursts of short duration), motivating a choice of bandwidth and time window similar in sizes or slightly larger than emission times for long duration gamma-ray bursts (100\,s and 600\,s, respectively). In essence, the bandwidth is designed to identify instantaneous times of interest for potential transient neutrino emission(s), while the time window is designed to collect all signal neutrinos localized in time around the potential transient phenomenon while keeping the number of background atmospheric neutrinos sufficiently low.

The background rate is not constant as a function of the azimuth angle due to the detector geometry having slightly different atmospheric muon rejection efficiency and neutrino acceptance, and is also not constant in time as a result of seasonal temperature variations in the atmosphere. In order to account for the fluctuations in time and azimuth, the discovery potential and sensitivity are estimated with an increased background rate of 15\%, by randomly duplicating 15\% of the events and adding them to the scrambled background. The result of an increase in the background is shown in Figure~\ref{fig: Sen_dis_15}, where the discovery potential is worsened by less than 5\%. Likewise, a decrease of 15\% in the background rate yields a less than 5\% better discovery potential.

\section{Results}
\label{sec:ResultsConlusion}
The unblinded (unscrambled) results based on data from April 2012 and May 2015 as well as the contribution of detector related systematic uncertainties are presented in this section.

\subsection{Detector systematic uncertainties}
\label{sec:Uncertainties}

We have estimated the detector related systematic uncertainties on the sensitivities by considering simulated data based upon different ice models~\citep{Aartsen:2013rt} and DOM efficiency~\citep{Aartsen:2016nxy}, i.e., the optical efficiency for a DOM to detect a photon. In these calculations, the most conservative ice models, i.e., the models resulting in the largest possible uncertainties, have been used. 

The ensuing systematic uncertainties on the sensitivity for a source following a subphotospheric energy spectrum with a mean energy of 100\,GeV are just above 25\% for downgoing neutrinos and around 17\% for upgoing events. The largest contribution arises due to the uncertainty on the model describing the bulk ice, e.g., different assumptions for the photon scattering and absorption within the bulk ice. 

The systematic uncertainties for the spectrum with mean energy 20\,GeV are slightly smaller, $\sim$20\% for downgoing and $\sim$15\% for upgoing neutrino events. In this case, the uncertainty on the DOM efficiency is the dominant factor.

\subsection{Unscrambled results}
\label{sec:UnblindResults}

\begin{figure}[t]
	\center
	\includegraphics[width=0.7\linewidth]{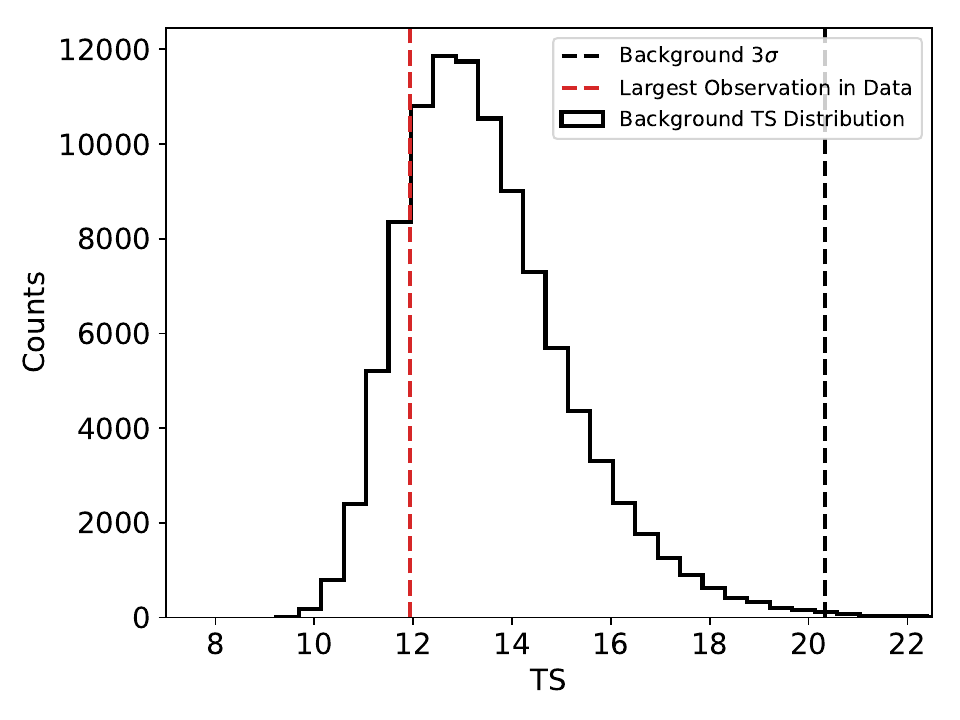}
	\caption{Test statistic value for the best-fit signal and source position (red dashed line) in the three years of GRECO data (2012-2015) compared to the nominal background TS distribution (without the addition of any systematic uncertainties) based on one year of data.}
	\label{fig: Result}
\end{figure}

When unscrambling the three years of GRECO data we find 300 cluster centers above the $R_{KDE}$-threshold resulting in 300 time windows. After applying the $n_s$-threshold, 267 time windows remain. The comparison between the TS for the best-fit signal and source position and the TS distribution for one year of scrambled background data is shown in Figure~\ref{fig: Result}. The 3$\sigma$ TS value for the background distribution is indicated by the black dashed line. The best-fit TS value is smaller than the 3$\sigma$ TS for background, and hence, this analysis reveals no transient neutrino emission from point sources in the data.

Since no transient neutrino emission from point sources was found in the data, we place 90\% C.L.\@ upper limits on the volumetric rate ($\dot{\rho}$) of the transient phenomena in the local universe. Because the observed TS from the data is an under fluctuation with respect to the expected median value derived from the scrambled trials (see Fig.~\ref{fig:TS_dists}), our upper limits are conservatively placed at the level of expected sensitivity. The discovery potential and sensitivity are given in terms of peak fluence, which is given by
\begin{linenomath}
\begin{equation} \label{eq: Peak fluence}
E^2 \Phi_\nu = \int_0^\infty \mathrm{d}z \frac{\mathcal{E}}{4\pi d_L^2(z)} \frac{\mathrm{d}N}{\mathrm{d}z} e^{-N(z)},
\end{equation}
\end{linenomath}
where the parameter $\mathcal{E}$ is the bolometric energy of the source and $N(z)$ describes the number of expected sources within redshift $z$ and is parameterized as
\begin{linenomath}
\begin{equation} \label{eq: Number_src}
N(z) = \dot{\rho} T f_{sky} \frac{4\pi}{3} d^3_c(z),
\end{equation}
\end{linenomath}
with $T$ being the livetime of the data used and $f_{sky}$ denotes the fraction of the sky considered. Since this work is a full sky search,  $f_{sky}=$1.
The factor 
$\frac{\mathrm{d}N}{\mathrm{d}z} e^{-N(z)}$
in Equation~\ref{eq: Peak fluence} is a measure of the probability of having the closest transient neutrino source at redshift $z$.

The relationship between the luminosity distance $d_L$ and the comoving distance $d_c$ is 
\begin{linenomath}
\begin{equation} \label{eq: Luminosity distance}
d_L(z) = (1+z)d_c(z),
\end{equation}
\end{linenomath}
where the comoving distance~\citep{Hogg:1999ad} can be expressed as
\begin{linenomath}
\begin{equation} \label{eq: comoving distance}
d_c(z) = \int_0^z \frac{c}{H(z')} \mathrm{d}z',
\end{equation}
\end{linenomath}
with $c$ being the speed of light and $H(z)$ is the Hubble parameter that takes the form 
\begin{linenomath}
\begin{equation} \label{eq: Hubble parameter}
H(z)^2 = H_0^2 \left( (1+z)^3 \Omega_m + \Omega_\Lambda \right),
\end{equation}
\end{linenomath}
by assuming a matter and dark energy dominated universe without curvature.
In this work, we adopt the value \mbox{74.03 $\pm$ 1.42\,$\text{km\,s}^{-1}\,\text{Mpc}^{-1}$} for the Hubble expansion rate based on the recent local universe measurements~\citep{Riess:2019cxk}. 
The cosmological parameters are taken to be $\Omega_M =$ 0.3 and $\Omega_\Lambda =$ 0.7, respectively.

The probability distribution describing the closest transient neutrino source with respect to redshift is then found by combining Equations~\ref{eq: Number_src} and \ref{eq: comoving distance} and differentiating with respect to the redshift. This yields
\begin{linenomath}
\begin{equation} \label{eq: dNdz}
\frac{\mathrm{d}N}{\mathrm{d}z} = \dot{\rho} T 4\pi \frac{c}{H(z)}  d_c^2.
\end{equation}
\end{linenomath}
By combining Equations~ \ref{eq: Peak fluence}, \ref{eq: Luminosity distance}, and \ref{eq: dNdz} we obtain
\begin{linenomath}
\begin{equation} \label{eq: Peak fluence 2}
\begin{split}
E^2 \Phi_\nu &= \int_0^\infty \mathrm{d}z \frac{\mathcal{E}}{4\pi (1+z)^2 d_c^2(z)} \dot{\rho} T 4\pi \frac{c}{H(z)} d_c^2(z) e^{-N(z)}, \\  
&= \dot{\rho} T c \mathcal{E} \int_0^\infty \mathrm{d}z \frac{1}{(1+z)^2 H(z)} e^{-N(z)}.
\end{split}
\end{equation}
\end{linenomath}
Finally, by using Equations~\ref{eq: Number_src}, \ref{eq: comoving distance}, and \ref{eq: Peak fluence 2} we get
\begin{linenomath}
\begin{equation} \label{eq: limit equation}
E^2 \Phi_\nu = \dot{\rho} T c \mathcal{E} \int_0^\infty \mathrm{d}z \frac{1}{(1+z)^2 H(z)} \exp \left( -\dot{\rho} T \frac{4\pi}{3} \int_0^z \frac{c}{H(z')} \mathrm{d}z'^3 \right).
\end{equation}
\end{linenomath}

\begin{figure*}
	\center
	\includegraphics[width=0.49\linewidth]{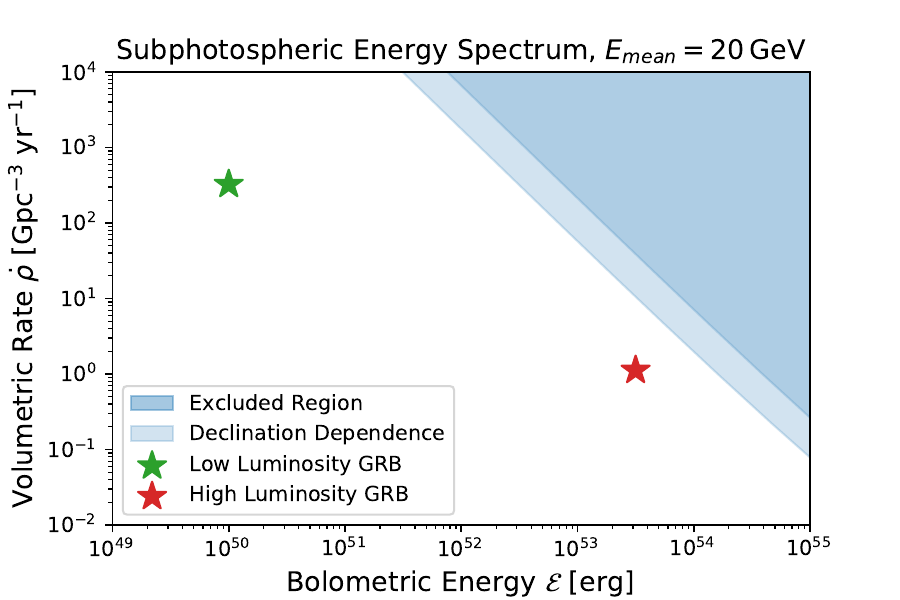}
	\includegraphics[width=0.49\linewidth]{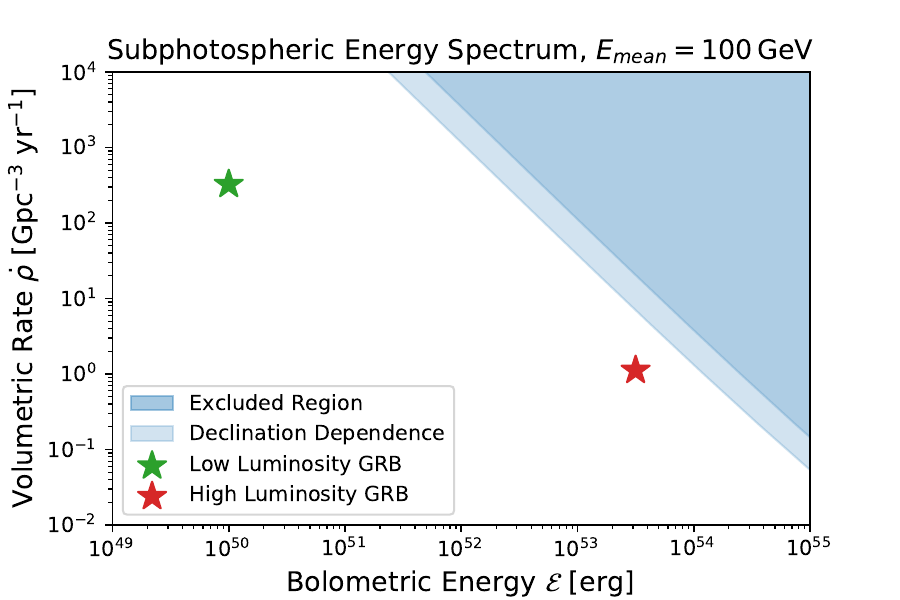}
	\caption{Upper limits (at 90\% C.L.\@) on the volumetric rate of transient neutrino point sources as a function of their bolometric neutrino energy. This is compared to models for high and low luminosity gamma-ray bursts, see~\citep{Murase:2013hh,Liang:2006ci}. The light blue band shows the declination dependence of the discovery potential. Left panel shows result based on sources with a mean energy of 20\,GeV, while the right panel is based on a mean energy of 100\,GeV. Both plots include the combined impact of a 15\% overall and 15\%-25\% declination dependent systematic uncertainty.}
	\label{fig: Limits}
\end{figure*}

From Equation~\ref{eq: limit equation} correlated values of the volumetric rate and the bolometric energy are found. The result is shown in Figure~\ref{fig: Limits}, where the 90\% upper limits include a rate increase of 15\% in order to account for the systematic uncertainties arising when assuming a constant background rate in time and azimuth. Likewise, the systematic uncertainties arising from assumptions for photon scattering, absorption within the bulk ice as well as uncertainties in the DOM efficiency reported in Section~\ref{sec:Uncertainties} have been accounted for in the final result. The upper limits use the sensitivities of the analysis derived from background-only scrambled trials, because the largest value of the test statistic based on data, reported in figure \ref{fig: Result}, is an under fluctuation. The sensitivity in terms of fluence varies with the declination of the source as seen in Figure~\ref{fig: Sen_dis}, and consequently the volumetric rate is declination dependent as well.

\section{Conclusion and Outlook}
\label{sec:Outlook}

This analysis has been performed on three years of GRECO data as an offline astrophysical search. This is the first all-flavor low-energy transient source search from IceCube and no astrophysical point sources were found. Upper limits of $\sim 705-2301\, \text{Gpc}^{-3}\, \text{yr}^{-1}$ have been placed on the volumetric rate of the transient neutrino sources, assuming neutrino spectra consistent with that from subphotospheric emission with flare times of up to approximately 600\,s, a mean neutrino energy of 100\,GeV, and a bolometric energy of $10^{52}$\,ergs. Effectively, this applies to gamma-ray bursts following the criteria described. Systematic uncertainties have been included, by using sensitivities calculated based on varying parameters as described above, ultimately making the limit more conservative. 

The analysis presented here is a potentially powerful tool for a real-time alert system. The ability to complement/extend the existing high-energy real-time alerts from IceCube, down to $\sim 5$\,GeV with an all-flavor neutrino selection opens up a new and relatively unexplored energy regime for the contribution of neutrinos to multimessenger astronomy. Future online as well as offline analyses will benefit from improvements in the angular reconstruction algorithms for low-energy neutrino events as well as improved multi-PMT photon sensors which will be deployed within IceCube as part of the IceCube Upgrade~\citep{Ishihara:2019aao} construction in the early to mid 2020s. A multimessenger campaign centered around an alert system with such improvements may render neutrino emission from high luminosity GRBs identifiable (see Figure \ref{fig: Limits}) in the near future.

\acknowledgments
The IceCube collaboration acknowledges the support from the following: USA {\textendash} U.S. National Science Foundation-Office of Polar Programs,
U.S. National Science Foundation-Physics Division,
Wisconsin Alumni Research Foundation,
Center for High Throughput Computing (CHTC) at the University of Wisconsin{\textendash}Madison,
Open Science Grid (OSG),
Extreme Science and Engineering Discovery Environment (XSEDE),
Frontera computing project at the Texas Advanced Computing Center,
U.S. Department of Energy-National Energy Research Scientific Computing Center,
Particle astrophysics research computing center at the University of Maryland,
Institute for Cyber-Enabled Research at Michigan State University,
and Astroparticle physics computational facility at Marquette University;
Belgium {\textendash} Funds for Scientific Research (FRS-FNRS and FWO),
FWO Odysseus and Big Science programmes,
and Belgian Federal Science Policy Office (Belspo);
Germany {\textendash} Bundesministerium f{\"u}r Bildung und Forschung (BMBF),
Deutsche Forschungsgemeinschaft (DFG),
Helmholtz Alliance for Astroparticle Physics (HAP),
Initiative and Networking Fund of the Helmholtz Association,
Deutsches Elektronen Synchrotron (DESY),
and High Performance Computing cluster of the RWTH Aachen;
Sweden {\textendash} Swedish Research Council,
Swedish Polar Research Secretariat,
Swedish National Infrastructure for Computing (SNIC),
and Knut and Alice Wallenberg Foundation;
Australia {\textendash} Australian Research Council;
Canada {\textendash} Natural Sciences and Engineering Research Council of Canada,
Calcul Qu{\'e}bec, Compute Ontario, Canada Foundation for Innovation, WestGrid, and Compute Canada;
Denmark {\textendash} Villum Fonden, Carlsberg Foundation;
New Zealand {\textendash} Marsden Fund;
Japan {\textendash} Japan Society for Promotion of Science (JSPS)
and Institute for Global Prominent Research (IGPR) of Chiba University;
Korea {\textendash} National Research Foundation of Korea (NRF);
Switzerland {\textendash} Swiss National Science Foundation (SNSF);
United Kingdom {\textendash} Department of Physics, University of Oxford.

\bibliographystyle{jhep}
\bibliography{references.bib}



\end{document}

%% file: Authors.tex
\author[16]{R. Abbasi,}
\author[56]{M. Ackermann,}
\author[17]{J. Adams,}
\author[11]{J. A. Aguilar,}
\author[21]{M. Ahlers,}
\author[47]{M. Ahrens,}
\author[27]{C. Alispach,}
\author[30]{A. A. Alves Jr.,}
\author[40]{N. M. Amin,}
\author[38]{K. Andeen,}

% ...
\affiliation[11]{Universit{\'e} Libre de Bruxelles, Science Faculty CP230, B-1050 Brussels, Belgium}
% ...
\affiliation[16]{Department of Physics, Loyola University Chicago, Chicago, IL 60660, USA}
\affiliation[17]{Dept. of Physics and Astronomy, University of Canterbury, Private Bag 4800, Christchurch, New Zealand}
% ...
\affiliation[21]{Niels Bohr Institute, University of Copenhagen, DK-2100 Copenhagen, Denmark}
% ...
\affiliation[27]{D{\'e}partement de physique nucl{\'e}aire et corpusculaire, Universit{\'e} de Gen{\`e}ve, CH-1211 Gen{\`e}ve, Switzerland}
% ...
\affiliation[30]{Karlsruhe Institute of Technology, Institute for Astroparticle Physics, D-76021 Karlsruhe, Germany }
% ...
\affiliation[38]{Department of Physics, Marquette University, Milwaukee, WI, 53201, USA}
% ...
\affiliation[40]{Bartol Research Institute and Dept. of Physics and Astronomy, University of Delaware, Newark, DE 19716, USA}
% ...
\affiliation[47]{Oskar Klein Centre and Dept. of Physics, Stockholm University, SE-10691 Stockholm, Sweden}
% ...
\affiliation[56]{DESY, D-15738 Zeuthen, Germany}
% ...
